\documentclass[amsmath,amssymb,aps,prl,superscriptaddress,twocolumn]{revtex4-2}
\usepackage{braket}
\usepackage{graphicx}% Include figure files
\usepackage{dcolumn}% Align table columns on decimal point
\usepackage{bm}% bold math
\usepackage{hyperref}
\usepackage{color}

\begin{document}
	
	\preprint{APS/123-QED}
	
	\title{Fractional Chern insulators in moir\'e flat bands with high Chern numbers}% Force line breaks with \\
 
\author{Chonghao Wang}
\thanks{These authors contributed equally to the work.}
\affiliation{State Key Laboratory of Low Dimensional Quantum Physics and
Department of Physics, Tsinghua University, Beijing, 100084, China}

\author{Xiaoyang Shen}
\thanks{These authors contributed equally to the work.}
\affiliation{State Key Laboratory of Low Dimensional Quantum Physics and
Department of Physics, Tsinghua University, Beijing, 100084, China} 

\author{Ruiping Guo}         
\thanks{These authors contributed equally to the work.}
\affiliation{State Key Laboratory of Low Dimensional Quantum Physics and
Department of Physics, Tsinghua University, Beijing, 100084, China}
\affiliation{Institute for Advanced Study, Tsinghua University, Beijing 100084, China}

\author{Chong Wang}
\affiliation{State Key Laboratory of Low Dimensional Quantum Physics and
Department of Physics, Tsinghua University, Beijing, 100084, China}

\author{Wenhui Duan}
\email{duanw@tsinghua.edu.cn}
\affiliation{State Key Laboratory of Low Dimensional Quantum Physics and
Department of Physics, Tsinghua University, Beijing, 100084, China}
\affiliation{Institute for Advanced Study, Tsinghua University, Beijing 100084, China}
\affiliation{Frontier Science Center for Quantum Information, Beijing, China}
\affiliation{Beijing Academy of Quantum Information Sciences, Beijing 100193, China}

\author{Yong Xu}
\email{yongxu@mail.tsinghua.edu.cn}
\thanks{}
\affiliation{State Key Laboratory of Low Dimensional Quantum Physics and
Department of Physics, Tsinghua University, Beijing, 100084, China}
\affiliation{Frontier Science Center for Quantum Information, Beijing, China}
\affiliation{RIKEN Center for Emergent Matter Science (CEMS), Wako, Saitama 351-0198, Japan}

\date{\today} 
	
	\begin{abstract}
		%In this study, we introduce a skyrmion potential to a $\Gamma$-valley twisted homobilayer transition metal dichalcogenide, resulting in the emergence of a flat Chern band with Chern number $C=-2$. \underline{The electron distribution in real space for this band exhibits a kagome lattice structure.} Subsequently, we observe the fractional Chern insulator phases in this high Chern band at various filling factors corresponding to the generalized Jain series $\nu=r/(2r\left| C\right|+1 )$ for $r=-2,-1,1$ cases. The total momentum of the degenerate ground states are all consistent with the generalized Pauli principle. Moreover, our calculation reveals that fractional Chern insulators can be realized in an experimentally achievable magnitude of exchange coupling strength, provided that the twist angle is as small as $\theta=2^{\circ}$.
        Recent discoveries of zero-field fractional Chern insulators in moir\'e materials have attracted intensive research interests. However, most current theoretical and experimental attempts focus on systems with low Chern number bands, in analogy to the Landau levels. Here we propose candidate material systems for realizing fractional Chern insulators with higher Chern numbers. The material setup involves $\Gamma$-valley twisted homobilayer transition metal dichalcogenides in proximity to a skyrmion lattice. The skyrmion exchange potential induces a flat band with a high Chern number $C = -2$. Using the momentum-space projected exact diagonalization method, we perform a comprehensive study at various filling factors, confirming the generalized Jain series. Our research provides theoretical guidance on realizing unconventional fractional Chern insulators beyond the Landau level picture.
	\end{abstract}
	
	\maketitle
	
	%\tableofcontents
	%\emph{Introduction.}--- In the past few decades, the fractional quantum Hall effect (FQHE) \cite{tsui1982two}, with its anyon excitations and potential applications in quantum computation, has attracted significant attention and interest. However, the requirement for strong magnetic fields significantly limits its technological applicability. Through advancements in the understanding of material magnetism and spin-orbit coupling, it has been revealed that these effects can be realized even without the necessity of magnetic fields in a topological flat band of lattice systems, named fractional Chern insulators (FCIs). \blue{maybe we can start directly from FCI without introducing the FQH.} 
    
 \emph{Introduction.}---The discovery of fractional quantum Hall effect (FQHE) have profoundly reshaped the landscape of condensed matter physics \cite{tsui1982two}. Following the discovery of quantum anomalous Hall effect, the zero-field counterpart of FQHE --- fractional Chern insulators (FCIs) --- has been theoretically proposed in toy tight-binding models \cite{wu2012zoology,regnault2011fractional,sheng2011fractional,neupert2011fractional,wang2011fractional,liu2013fractional,kourtis2014fractional}. These tight-binding models feature flat Chern bands, and host FCI states at certain fractional fillings of the flat Chern bands. While fine-tuned toy models are unlikely to materialize in real materials, recent breakthrough experiments show that moir\'e superlattices are perfect platforms to realize flat Chern bands, and in turn zero-field fractional Chern insulators \cite{cai2023signatures,park2023observation,xu2023observation,kang2024evidence,zeng2023thermodynamic,lu2024fractional,xie2024even}. These experiments have spurred a flurry of theoretical and numerical studies to model the relevant moir\'e superlattices.

The FCI states experimentally observed in moir\'e superlattices have been found within flat bands with Chern numbers $\left| C\right| =1$. These flat bands can be continuously deformed to the Landau level \cite{moller2015fractional}. Remarkably, topological bands in lattices can host flat bands with high Chern numbers $\left| C \right| > 1$. FCI states in these high-Chern-number flat bands (designated as HC-FCI) do not find analogue in Landau level systems and have been proposed in toy tight-binding models and Hofstadter models \cite{moller2015fractional,andrews2021stability,liu2012fractional,wang2012fractional,sterdyniak2013series}. Given the success of observing $\left| C \right| = 1$ FCI states in moi\'e superlattices, it is reasonable to anticipate that HC-FCI states will be more readily realized in these systems. Indeed, HC-FCI has been proposed in twisted multilayer graphene superlattices, where the high-Chern number flat bands can be revealed if the four-fold spin-and-valley degeneracy is lifted \cite{wang2022hierarchy,dong2023many}. In the $\Gamma$-valley transition metal dichalcogenide (TMD) moir\'e superlattices, the high-Chern-number flat Chern bands can be exposed by only lifting the two-fold spin degeneracy, presenting an advantage in engineering high-Chern-number flat bands in these systems.%In the transition metal dichalcogenide (TMD) moir\'e superlattices, due to the spin-valley locking, the high-Chern-number flat Chern bands can be exposed by only lifting the two-fold spin-or-valley degeneracy, presenting an advantage in engineering high-Chern-number flat bands in these systems.

%enabling them to host the FCI, which surpasses the Landau level picture \cite{moller2015fractional}.   

%Remarkably, topological bands in lattices can exhibit Chern numbers $\left| C\right| >1$, enabling them to host the FCI, which surpasses the Landau level picture \cite{moller2015fractional}.  Experimental observations have revealed FCI for high Chern band in van der Waals heterostructures under an external magnetic field \cite{spanton2018observation}. Several numerical calculations have been done to investigate the FCI at $1/5$ filling with Chern number $\left| C\right| =2$ in Harper-Hofstadter bands  \cite{andrews2021stability}, as well as the FCI in multilayer kagome lattices \cite{liu2012fractional} and three-band triangular lattices\cite{wang2012fractional}. Additionally, in twisted bilayers of Bernal-stacked multilayer graphene, FCI in high Chern bands have been calculated using ED approach \cite{wang2022hierarchy,dong2023many}. It is crucial to search for more achievable materials capable of supporting the FCI with higher Chern number. This pursuit may help us to establish a more general theory on FCI for there is no Landau level corrspondence and may even realize the non-Abelian FCIs within $\left| C\right| >1$  topological bands \cite{sterdyniak2013series}.
	
In this work, we propose a potential platform to realize HC-FCI based on TMD moir\'e superlattices. By introducing a skyrmion potential onto the $\Gamma$-valley twisted homobilayer TMD systems, we find that the third moir\'e valence band is flat, with Chern number $C=-2$. %$\mp 2$ for the $K$ and $-K$ valley. 
The skyrmion potential might be realized by magnetic proximity effects between the TMD moir\'e superlattices and twisted 2D magnets. Using the momentum space projected exact diagonalization method, we confirm the existence of HC-FCI in a generalized Jain series at hole fillings $\nu = -2/7,-1/3,-1/5,-2/9$. Our research provides theoretical guidance on realizing unconventional fractional Chern insulators beyond the Landau level picture.

%which can be provided by substrates such as twisted $\rm{CrI_3}$ or $\rm{CrBr_3}$~\cite{li2023deep}. Under the skyrmion potential, moir\'e bands become more flatter and the third valence band obtains a Chern numer $C=-2$. In 2015, Gunnar M\"ller and Nigel Cooper propose FCIs in a series of different fillings described by generalized Jain series $\nu=r/\left( 2r\left| C\right|+1 \right) $\cite{moller2015fractional}. Here, we confirm this series by ED calculation on this flat band. The results indicate that FCI phase emerges as the many-body ground state at $r=-2,-1,1,2$ corresponding to the fillings of $\nu=-2/7,-1/3,-1/5,-2/9$, where the minus sign represents the hole doping.

    \begin{figure*}[t]
			\centering
			\includegraphics[width = 0.98\textwidth]{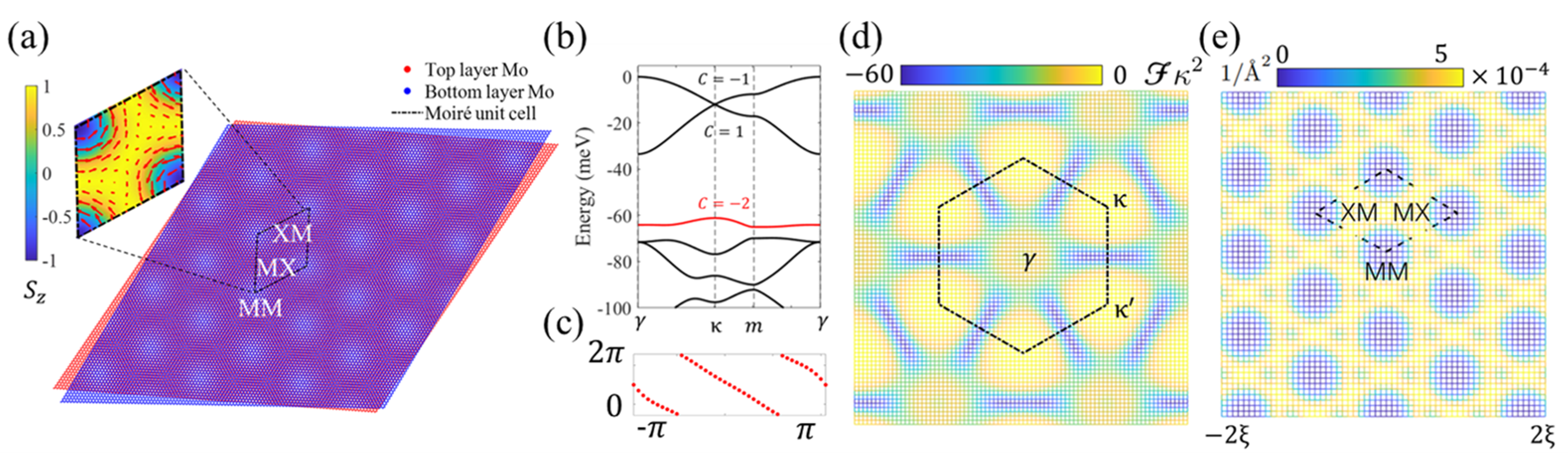}
			\caption{\label{BandStructure} (a) Real space image of twisted homobilayer TMD with the skyrmion potential.  (b) Bloch band structure in twisted $\rm{MoS_2}$ with exchange coupling strength $J=40$ meV and twist angle $\theta=3.15^{\circ}$.The Chern numbers for the first three bands are $-1$, $1$, $-2$, respectively. (c),(d) are the Wilson loop and Berry curvature distribution for the third band. $\gamma$, $\kappa$, $\kappa^{\prime}$ are the three high symmetry k-points in the moir\'e Brillouin zone. (e) Charge distribution for the third band in real space, which exhibits a kagome lattice-like pattern. }
	\end{figure*}   

\emph{Continuum model.}---We start with the moir\'e Hamiltonian of $\Gamma$-valley twisted bilayer TMD in the presence of a magnet substrate carrying a periodic skyrmion texture $S(\mathbf{r})$~\cite{angeli2021gamma,shen2024stabilizing}. The exchange coupling between the twisted bilayer TMD and the periodic skyrmion lattice is included and its strength is represented by $J$. Due to interlayer hybridization, the antibonding state at the $\Gamma$-valley has a gap of over 350 meV with other bands. By focusing solely on this antibonding state, the model Hamiltonian can be simplified \cite{angeli2021gamma}.
\begin{eqnarray}
    H_0=\left[ -\frac{\hbar^2 k^2}{2m^{\star}} + \Delta\left( \mathbf{r}\right) \right] \sigma_0+J \vec{S}\left( \mathbf{r}\right)  \cdot \vec{\sigma}	
\end{eqnarray}
Here, $\sigma$ represents the spin of the valence band electrons. $\Delta{\left( \mathbf{r}\right) }$is the moir\'e potential, characterized by moir\'e lattice periodicity, $m^\star$ is the effective mass of electrons in the valence band. It satisfies the following expression under the consideration of symmetry:
	
	\begin{eqnarray}
		\Delta\left( \mathbf{r}\right)=\sum_{s=1}^3 \sum_{j=1}^{6} V_s \exp\left(i \mathbf{g}_j^s \cdot r + i \phi_s \right)  
	\end{eqnarray} 
	The superscript $s$ refers to the s-th shell of continuum model basis. The parameters of the continuum model in Fig.\ref{BandStructure}.(a) are determined by fitting the band structure calculated from large-scale density functional theory (DFT) simulations\cite{angeli2021gamma}. The units of $V_1$, $V_2$, and $V_3$ in TABLE I. are millielectronvolts (meV), while the effective mass $m^{\star}$ is expressed in units of the bare electron mass $m_e$. Additionally, the lattice constants of monolayer TMDs are measured in Angstroms \cite{angeli2021gamma}.
	
	\begin{table}[htp]
		\begin{ruledtabular}
			\caption{parameters for continuum model}
			\begin{tabular}{lcccccr}
				\textrm{Materials}&
				\textrm{$V_1$(meV)}&
				\textrm{$V_2$(meV)}&
				\textrm{$V_3$}(meV)&
				\textrm{$\phi_{1,2,3}$}&
				\textrm{$m^{\star}/m_e$}&
				\textrm{$a_0  (\text{\AA})$} \\
				\colrule
				$\textrm{MoS}_2$ & 39.45 & 6.5 & 10 & $\pi$ &0.9 &3.182 \\
				$\textrm{WS}_2$ & 33.5 & 4 & 5.5 & $\pi$ &0.87 &3.18\\
				$\textrm{MoSe}_2$ & 36.8 & 8.4 & 10.2 & $\pi$ &1.17 &3.295\\
			\end{tabular}
		\end{ruledtabular}
	\end{table}
	
	We assume that the skyrmion potential also shares the same periodicity with the twisted TMD. Skyrmion configurations with highly tunable periods have been proposed in several materials\cite{tong2018skyrmions,xie2023evidence,hejazi2020noncollinear,xu2022coexisting,song2021direct}.  And we prove that the HC-FCI can survive across various twist angles, allowing us to adjust the TMD twist angles to match the periodicity of the skyrmion potential. The generic skyrmion potential is expressed as $\vec{S}\left( \mathbf{r}\right)=\vec{N}\left( \mathbf{r}\right) /| \vec{N}\left( \mathbf{r}\right)| $, with $\vec{N}\left( \mathbf{r}\right)$ defined as\cite{paul2023giant}:
	
	\begin{eqnarray}
		\vec{N}\left( \mathbf{r} \right)= \frac{1}{\sqrt{2}} \sum_{j=1}^{6} e^{i \mathbf{g}_j^1 \cdot \mathbf{r}} \hat{e}_j+m_z \hat{e}_z
	\end{eqnarray} 
	where $\mathbf{g}_j^1=\mathbf{G}_j^{m}=4\pi (\cos\theta_j,\sin\theta_j)/\left( \sqrt{3}\xi\right) $ are moir\'e reciprocal lattice vectors and $\xi$ is the moir\'e wavelength. $\hat{e}_j=\left( i\sin\theta_j,-i\cos\theta_j,-1\right)/\sqrt{2} $ with these six angles are arranged in the following order: $\theta_j=\left( j-1\right) \pi/3$.

	\emph{Band structures.} --- To precisely capture the profile of the skyrmion texture, we truncate up to the forth harmonic terms in $\vec{S}(\mathbf{r})$. 
	
	\begin{eqnarray}
		\vec{S}(\mathbf{r})=\bar{m}_z \hat{e}_z+ \left( \sum_{s=1}^4 \sum_{j=1,3,5} \vec{T}_j^s e^{i\mathbf{g}_j^{s} \cdot \mathbf{r}} +h.c. \right) 
	\end{eqnarray}
	$T_j^s$ are numerically obtained via the Fourier transformation:
	
	\begin{eqnarray}
		\vec{T}_j^s=\frac{1}{\Omega} \int d \mathbf{r} \vec{S}(\mathbf{r}) e^{-i \mathbf{g}_j^s \cdot \mathbf{r}}
	\end{eqnarray}
	Here, $\Omega$ represents the volume of the system, and the term $\bar{m}_z \hat{e}_z$ denotes the average of $\hat{S}(\mathbf{r})$ within a moir\'e unit cell.  The parameter $\bar{m}_z=0.5$ is obtained by setting $m_z=1.16$, which dictates the magnitude of the average Zeeman splitting. In this context, the Zeeman splitting is equal to $J$.  Before introducing the skyrmion potential, the top two bands have a cross at $\kappa$ point, which is  protected by the $D_{3}$ point group, and can be gapped by an out of plane gating field\cite{zhang2021electronic}. In this condition, these two bands are topologically trivial and therefore do not raise interest. Remarkably, the skyrmion potential breaks both time-reversal symmetry (TRS) and $C_{2y}$ symmetry, and gap out both $\kappa$ and $\gamma$ points.
	
	Under the skyrmion potential, all bands in FIG.1.(a) are endowed with non-vanishing Chern numbers, as we have labeled the Chern numbers of the top three bands. We calculate the Wilson loop and Berry curvature of the third band, as shown in FIG.\ref{BandStructure}.(c) and (d), which clearly indicates the Chern number equals to -2. Subsequently, we also calculate the real space charge distribution shown in FIG.\ref{BandStructure}.(e), suggesting a potential connection with an effective kagome lattice \cite{liu2012fractional}. 
	
	\emph{Numerical results for HC-FCI} ---  Following the construction of the continuum model and determination of the Chern numbers, we proceed to compute the many-body ground state by introducing a dual-gate screened Coulomb interaction \cite{wang2024fractional}:
	
	\begin{eqnarray}
		H_{int}=\frac{1}{2\Omega} \sum_{\mathbf{k,p,q}} V(q) a_{\mathbf{k}+\mathbf{q}}^{\dagger} a_{\mathbf{p}-\mathbf{q}}^{\dagger} a_{\mathbf{p}} a_{\mathbf{k}} 
	\end{eqnarray}
    With $V(q)=\frac{\tanh(qd)}{2\varepsilon\varepsilon_0 q}$, where $d$ represents the gate distance, $\varepsilon$ is the relative dielectric constant, $\varepsilon_0$ denotes the permittivity of vacuum, and $a_{\mathbf{k}}$ represents the annihilation operator of the continuum model basis—a plane wave basis. Note that in the above formula, the summation is over the whole reciprocal space. We then project the interactions onto the third valence band, which has a Chern number $C = -2$ \cite{abouelkomsan2020particle}.
	
	\begin{eqnarray}
		H_{int}&=&\frac{1}{2\Omega} \sum_{\mathbf{G}} \sum_{\mathbf{k,p,q}}^{BZ} V(\mathbf{q}+\mathbf{G})  F_{\mathbf{k+q+G,k}} F_{\mathbf{p-q-G,p}} \nonumber \\ 
		&&   \times C_{\mathbf{k}+\mathbf{q}}^{\dagger} C_{\mathbf{p}-\mathbf{q}}^{\dagger} C_{\mathbf{p}} C_{\mathbf{k}} 
	\end{eqnarray}
	
	The momentum vectors $\mathbf{k}$, $\mathbf{p}$, and $\mathbf{q}$ in the equation above all belong to the first Brillouin zone (BZ). While the form factors $F_{\mathbf{k},\mathbf{k^{\prime}}}$ are defined as $\left\langle \varphi_\mathbf{k} | \varphi_{\mathbf{k^{\prime}}} \right\rangle $, where $ \ket{\varphi_{\mathbf{k}}} $ represents the eigenstate of the continuum model for the third band. It's worth noting that the state $\ket{\varphi_{\mathbf{k+G}}}$ can be obtained using the relation $\ket{\varphi_{\mathbf{k+G}}}=V^{\mathbf{G}}\ket{\varphi_{\mathbf{k}}}$, where $V^{\mathbf{G}}$ represents the embedding matrix. Here, $\mathbf{G}=m_1\mathbf{G_1}+m_2\mathbf{G_2}$ is an arbitrary reciprocal lattice vector, and a cutoff $ \left| m_{1,2} \right| \leq 3$ suffices to obtain accurate results, as the form factors decay rapidly with increasing $\left| \mathbf{G} \right| $.

		\begin{figure}[htp]
			\centering
			\includegraphics[width = 8.6cm]{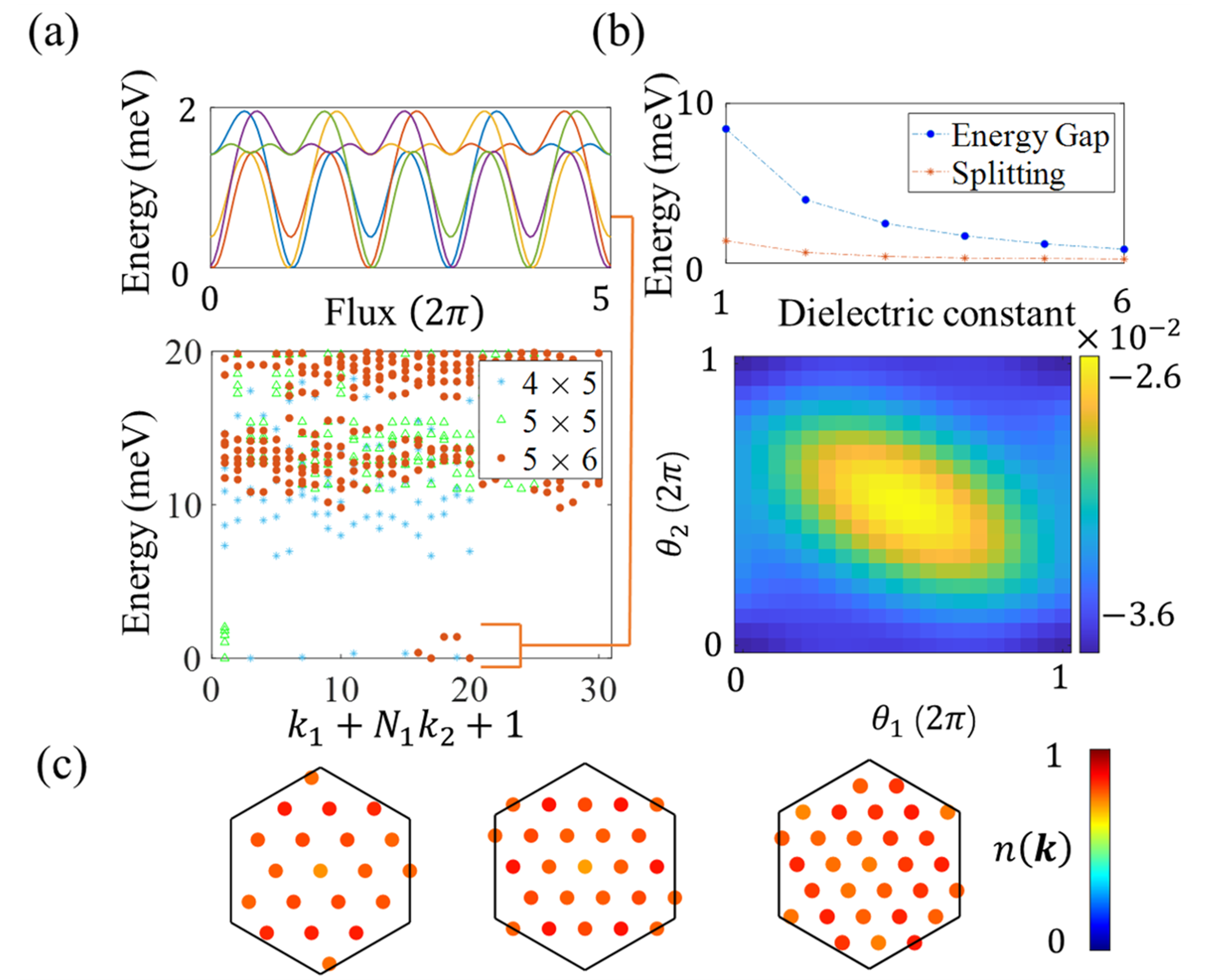}
			\caption{\label{ED_MoS2} (a) Many-body energy spectrum for -1/5 filling of the third band for different k-point numbers (20, 25 and 30 k-points) and energy spectral flow with the flux insertion along $G_6$ direction for 30-k-point situation. 5 different colors denote the five degenerate ground states. These five states mix with each other after a $2\pi$ flux insertion and return to their original sequence after $5 \time 2\pi$ flux insertion. (b) Upper panel shows the nany-body gap and splitting as a function of dielectric constant. The FCI phase is still observed when the relative dielectric constant $\epsilon$ is set to 6. Lower panel denotes Berry curvature distribution in the twisted phase space. The many-body Chern number equals to -2/5. (c) The electron distribution in BZ for cases of 20, 25, 30 k-points. The uniform distribution is one of the evidence for FCI phase. }
		\end{figure}

	The ED calculation is performed at a twist angle of $3.15^{\circ}$ with exchange coupling strength $J=40$ meV for twisted $\rm{MoS_2}$. According to the generalized Jain series $\nu=r/(2r\left| C\right|+1 ), \left( r \in \mathbb{Z}\verb|\| \{0\} \right) $, there should be five degenerate ground states for the $r=1$ case when the ground state is a FCI phase. For clarity, we denote the number of holes as $N_h$, and the total number of k-points as $N_k=N_1 \times N_2$. Thus, the filling factor can be defined as $\nu=-N_h/N_k$, where the minus sign indicates hole doping. We distribute these k-points uniformly in the BZ along the $\mathbf{G}_1$ direction and $\mathbf{G}_6$ direction, with $N_1 \times N_2$ k-points. We then perform the ED calculation with the total number of k-points set as $N_1 \times N_2=4 \times 5$, $5\times 5$, and $5\times 6$ at the fixed filling $\nu = -\frac{1}{5}$. The many-body energy spectrum for these scenarios are shown in FIG.\ref{ED_MoS2}.(a), where five nearly degenerate states are present in both three cases. According to the experiment\cite{tsui1982two} , FCI phase has a many-body gap with other many-body states. The gap between the first excited state and the five-fold degenerate ground states is defined as $\text{Gap}=E_6-E_5$, where $E_ j$ represents the $j^{th}$ lowest energy. And the splitting of the five degenerate ground states is defined as $E_5-E_1$. Due to the finite size effects, the gap of FCI increases with the number of k-points, which indicates that the system is an FCI in the thermodynamic limit.
	
	Apart from FCI, it is worth noting that one of the competing phases, the charge density wave (CDW), may also exhibit multiple degenerate ground states and a many-body gap between other states. To exclude the CDW phase, we verify our results using the generalized Pauli principle\cite{regnault2011fractional,bernevig2012emergent}, which determines the total momentum for the degenerate FCI ground states. Following the usual practice, we arrange all the k-points in one dimension within the range of $k=k_1+N_1k_2+1$ ($k_1=0,1,\cdots N_1-1$ and $k_2=0,1,\cdots N_2-1$ ) in FIG.\ref{ED_MoS2}.(a). In this sequence, the five degenerate FCI states should present at $k=3,7,11,15,19$ for $N_1\times N_2=4\times 5$ case, and $k=16,17,18,19,20$ for $N_1\times N_2=5\times6 $ case. As for $N_1 \times N_2=5\times 5$, with both $N_1$ and $N_2$ being multiples of 5, all 5 states have the same total momentum and fall on the same k-point, which here is the $\gamma$ point. In addition to the generalized Pauli principle, The flux insertion can also verify the FCI phase and remove the possibility of CDW. In FIG.\ref{ED_MoS2}.(a), we insert flux along the $\mathbf{G}_6$ direction for the case of $N_1 \times N_2=5 \times 6 $, and the energy spectral flow has a periodicity of $5 \times 2\pi$, indicating the many-body Chern number $C_{mb}=-2/5$. During the flux insertion, a global gap still exists. In FIG. \ref{ED_MoS2}.(b), we vary the dielectric constant and observe a decrease in the gap as the dielectric constant increases. In fact, the energy scale of the interaction exceeds the band gap between the third band and other bands in the continuum model, allowing the other bands to also influence the third band via the Coulomb interaction \cite{wang2024higher,xu2024maximally}. And we will consider this effect in a future study. Then we calculated the many-body Chern number of the ground states by introducing the twisted boundary condition\cite{kudo2019many}:
	\begin{eqnarray}
		\Psi\left( \mathbf{r}+N_{1,2}\mathbf{a}_{1,2}\right) =e^{i\theta_{1,2}}\Psi\left( \mathbf{r}\right)  
	\end{eqnarray}
	Here, $\theta_{1,2} \in \left[ 0,2\pi\right) $ is a constant and depends on the magnitude of the flux $\phi$. In the special case of $\phi=0$, the twisted boundary condition reduces to a periodic boundary condition. The Berry curvature distribution is quite uniform across the boundary phase space, which is consistent with previous theoretical predictions and numerical findings\cite{kudo2019many,niu1985quantized}. Notably, we have performed a particle-hole transformation and calculated the Chern numbers using the wave functions of holes. This results in the relationship between the many-body Chern number and hole fillings given by $C_{mb}=C \left| \nu \right|$. Finally, the uniform charge distribution in the BZ was examined. The electron charge numbers with total momentum $k$ can be calculated using the equation $n_k=\bra{\psi_{GS}} C_{k}^{\dagger} C_{k} \ket{\psi_{GS}}$. FIG. \ref{ED_MoS2}.(c) illustrates the charge distribution of the ground state, showing values close to 4/5 for electrons at each k-point.   
	
	\begin{figure}[h]
		\includegraphics[width=8.6cm]{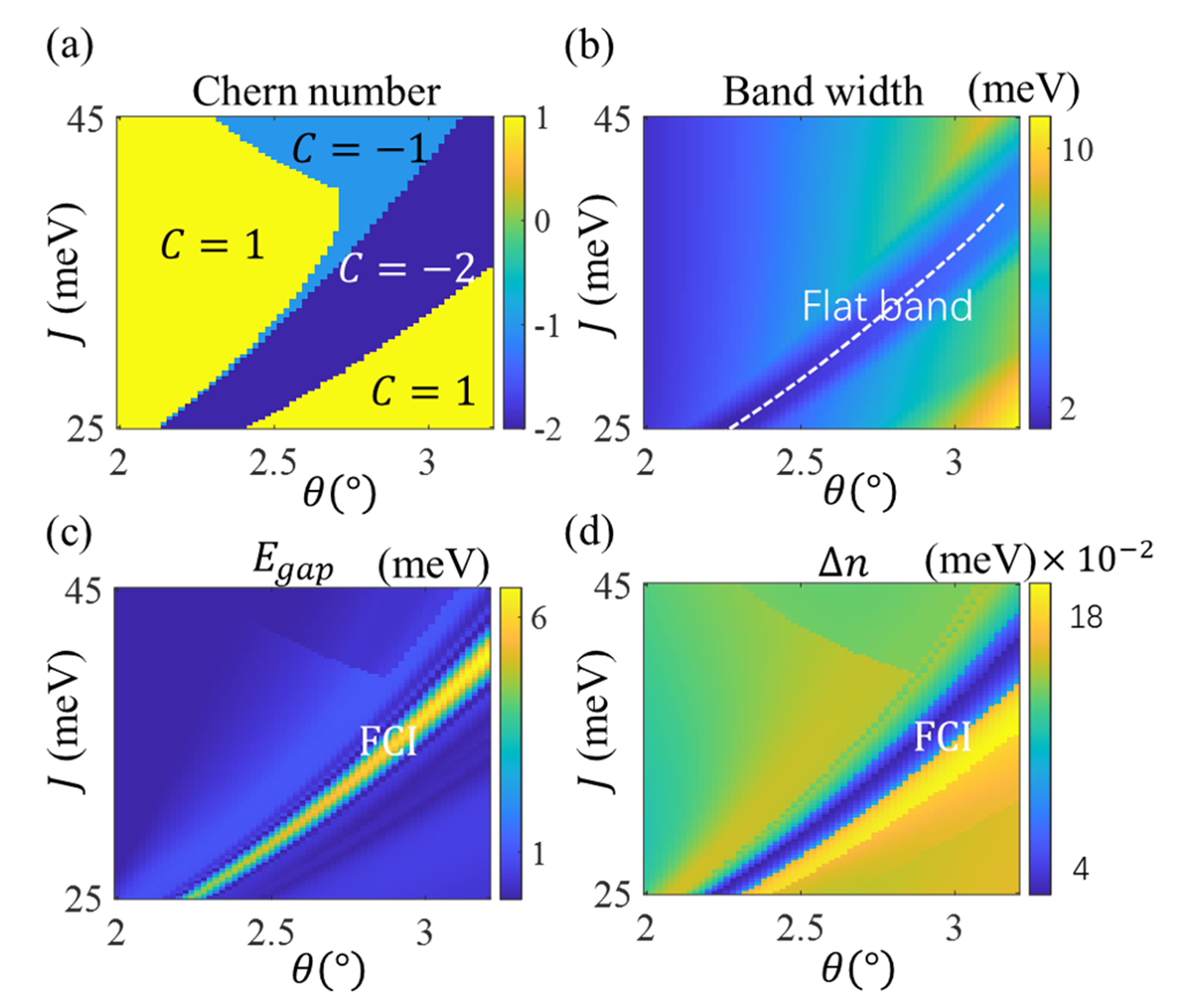}% Here is how to import EPS art
		\caption{\label{Twistangle} Phase diagrams for the third band. (a) shows the Chern number phase diagram for the third band as a function of twist angles and exchange energy. (b) After calculating the band width, we identify a flat band with Chern number $C=-2$ within the region marked by the white dashed line. (c) A high Chern number FCI region, characterized by the many-body energy gap $E_{gap}=E_6-E_5$ is highlighted in bright color. In the whole phase diagram, the k-points number is 20 and the filling factor is set to be $\nu=-1/5$. (d) The electron distribution is calculated to verify the FCI phase by $\Delta n$ defined as $\Delta n=\frac{1}{ N_{\mathbf{k}}} \sum_{ N_{\mathbf{k}}} \left|  n(\mathbf{k})-\bar{n}(\mathbf{k})  \right|  $, $n(\mathbf{k})$ being the electron number at $\mathbf{k}$ point and $\bar{n}$ is the averaging density over BZ., with $n(\mathbf{k})$ being the electron number at $\mathbf{k}$ point. And it reaches the smallest values in FCI region. }
	\end{figure}

    \emph{Magnitude of exchange coupling strength} --- According to the DFT simulations, a giant Zeeman splitting of approximately 17 meV or even 19 meV has been observed for $\rm{MoS_2}$ with twisted $\rm{CrBr_3}$ heterostructures\cite{paul2023giant}, suggesting a sizable order of exchange interaction strength $J$. So that another concern is how to decrease the magnitude of the exchange coupling strength needed to realize FCIs, especially considering that $J=40$ meV in FIG.\ref{ED_MoS2} is relatively large. Based on previous research\cite{paul2023giant}, a dimensionless quantity $\frac{Jm^{\star} \xi^2}{\hbar^2}$ can be formulated to compare the magnitude of the exchange coupling strength with the kinetic energy of electrons in moir\'e bands. This dimensionless quantity suggests that a smaller exchange coupling is required for larger moir\'e unit cells or smaller twist angles, in other words. To validate our viewpoint, we then sweep the phase diagram as functions of $\theta $ and $J$ in the physical range. We perform $4\times 5$ k points ED when sweeping the phase diagram. A smaller exchange coupling strength is required to realize the FCI, as shown in FIG.\ref{Twistangle}(c), and the same phenomenon also exists in twisted $\rm{MoSe_2}$ and $\rm{WS_2}$. As verified by ED calculation in twisted $\rm{MoSe_2}$, FCI can occur at $J=18$ meV at twist angles of 2.2 degree, which falls within the range of exchange coupling strengths in twisted $\rm{CrBr_3}$. 
   
      \begin{figure}[h]
		\includegraphics[width=8.6cm]{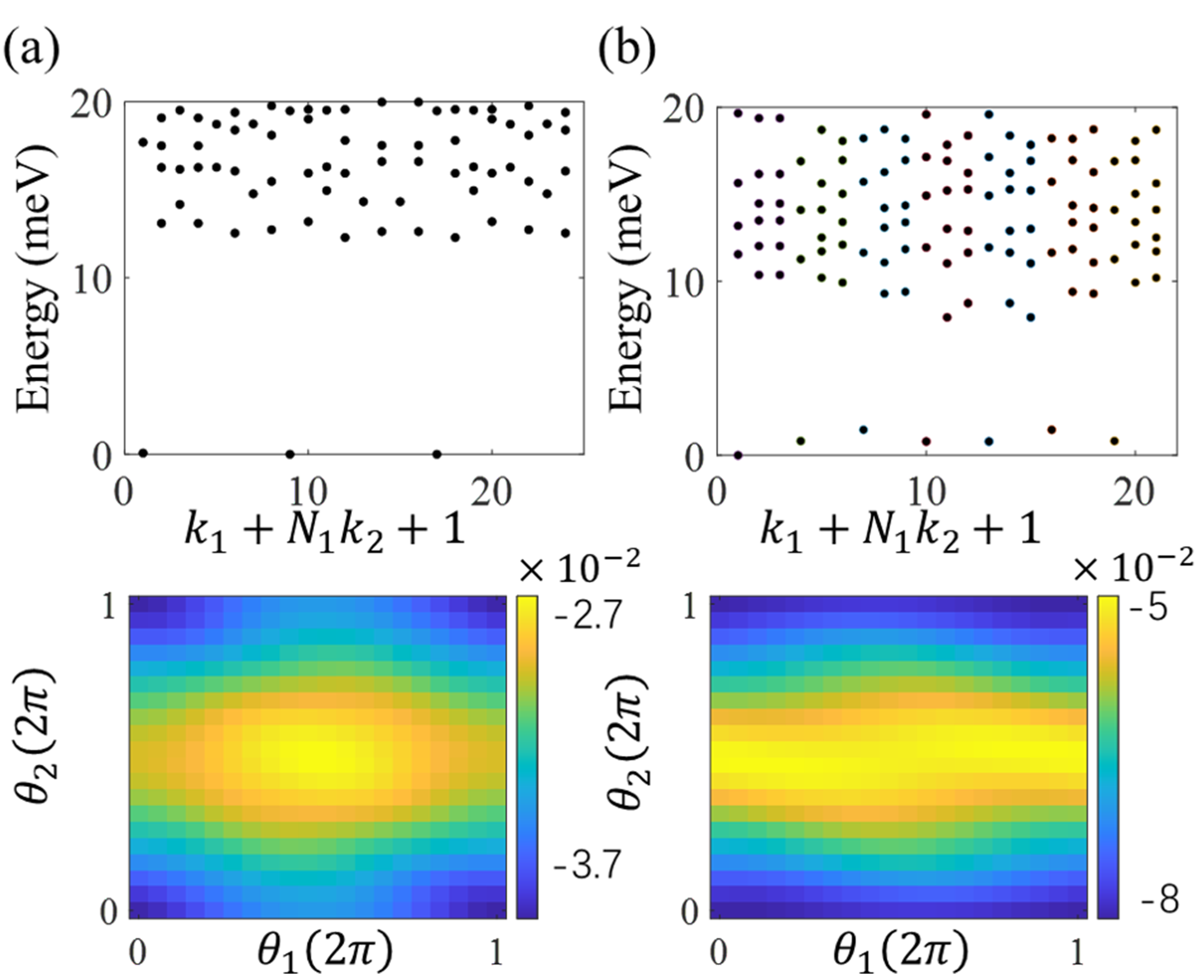}% Here is how to import EPS art
		\caption{\label{Otherfillings} Upper panel shows the many-body energy spectrum of the twisted $\rm{MoS_2}$ at $\theta=3.15^{\circ}$ obtained by ED calculation. The lower panel represents the Berry curvature for the ground state under the twisted boundary condition. (a). -1/3 filling in $N_1\times N_2=4\times6$ k-points and many-body Chern number equals to -2/3. (b) shows the energy spectrum of -2/7 filling in $N_1\times N_2=3\times9$ k-points with 7 degenerate ground states. And the many-body Chern number equals to -4/7. }
	\end{figure}

	\emph{Many-body Chern number in other fillings} ---Additionally, further ED calculations for various filling factors have been conducted based on the function $\nu=r/(2r\left| C\right|+1 ), \left( r \in \mathbb{Z}\backslash  \{0\} \right) $. By setting $r=-1,-2$, we obtain the filling factors $\nu=-1/3, -2/7$, where the negative sign denotes hole doping. We take the $\rm{MoS_2}$ as an example with twisted angle $\theta=3.15^{\circ}$. Similar results for $\rm{WS_2}$ and $\rm{MoSe_2}$ are also calculated in Supplemental Materials.  For $-1/3$ filling, we sample $N_1 \times N_2=4\times6$ k-points in the BZ, and observe three degenerate ground states at $k=1, 9, 17$ points which meets the generalized Pauli principle. And the electron distribution is also uniform and nearly equals to $2/3$ for each k-point. For a general filling ($\left| r \right|>1 $), denoted as $\nu=p/q$, there exist $q$ degenerate states, with no more than $p$ particles (in this case, holes) occupying $q$ consecutive orbitals\cite{bergholtz2008quantum}. The latter conclusion is just the generalized Pauli principle, which we used to check our results. These conclusion can be theoretically obtained through a thin torus limit (Tao-Thouless limit\cite{tao1983fractional}), where the many-body problem can be exactly solved. Based on these conclusions, we observed 7 nearly degenerate ground states at $k=1,4,7,10,13,16,19$ points, consistent with the generalized Pauli principle (further details are provided in the Supplemental Materials). And we also calculate the energy spectrum at fillings of -2/9, which is correspondent to the $r=2$ case in Jain series. Similarly, 9 nearly degenerate states with a small gap are found at $k=1,10,19,1,10,19,1,10,19$, also in accordance with the generalized Pauli principle. The ED results can be found in the Supplemental Materials. Moreover, the electron distribution is uniform for both $\nu=-2/7$ and $\nu=-2/9$, nearly equaling $5/7$ and $7/9$ at each $k$-point, respectively. And the Berry curvatures for the parameter space of $\vec{\theta}=\left( \theta_1,\theta_2 \right) $ of fillings $-1/3$ and $-2/7$ are shown in FIG.\ref{Otherfillings}(a) and (b).

	%Next, we calculate the many-body Chern number of the ground states by introducing the twisted boundary condition\cite{kudo2019many}:
	%\begin{eqnarray}
	%	\Psi\left( \mathbf{r}+N_{1,2}\mathbf{a}_{1,2}\right) =e^{i\theta_{1,2}}\Psi\left( %\mathbf{r}\right)  
	%\end{eqnarray}
	%Here, $\theta_{1,2} \in \left[ 0,2\pi\right) $ is a constant and depends on the magnitude of the flux $\phi$. In the special case of $\phi=0$, the twisted boundary condition reduces to a periodic boundary condition. The Berry curvatures for the parameter space of $\vec{\theta}=\left( \theta_1,\theta_2 \right) $ of fillings $-1/3$ and $-2/7$ are shown in FIG.\ref{Otherfillings}(a) and (b). The Berry curvature distribution is quite uniform across the boundary phase space, which is consistent with previous theoretical predictions and numerical findings\cite{kudo2019many,niu1985quantized}. Notably, we have performed a particle-hole transformation and calculated the Chern numbers for these fillings using the wave functions of holes. This results in the relationship between the many-body Chern number and hole fillings given by $C_{mb}=C \left| \nu \right|$.

	\emph{Conclusion.} --- In this study, we investigate the fraction Chern insulator with 
 higher chern number in twisted TMD under the assist of the skyrmion lattice.  We have broken the band degeneracy in the $\Gamma$ valley twisted homobilayer TMD, generating nontrivial moir\'e flat bands by introducing a skyrmion potential. Additionally, under certain appropriate values of the exchange coupling strength, a flat band with a higher Chern number $C=-2$ emerges. The real space electron distribution for this high Chern band exhibits characteristics akin to an effective kagome lattice. Subsequently, through k-space projected ED calculations, we observe the features of FCI at various filling factors, including $\nu=-1/5$, $-1/3$, $-2/7$ and $-2/9$. The observed degenerate ground states numbers, their total momentum and also the uniform charge distribution all align with theoretical expectations, which verify the existence of FCI in high Chern band. Notably, a filling factor of $\nu=-4/5$ was also examined, revealing a smaller FCI gap and a larger ground state splitting. In addition to the two-band model, we also developed a four-band continuum model for twisted $\rm{MoS_2}$ with a skyrmion potential, where all the aforementioned conclusions can be observed. The numerical results are provided in the Supplemental Materials.

	\appendix

	\bibliographystyle{unsrt}
	\bibliography{reference.bib}

\begin{thebibliography}{10}

\bibitem{tsui1982two}
Daniel~C Tsui, Horst~L Stormer, and Arthur~C Gossard.
\newblock Two-dimensional magnetotransport in the extreme quantum limit.
\newblock {\em Physical Review Letters}, 48(22):1559, 1982.

\bibitem{wu2012zoology}
Yang-Le Wu, B~Andrei Bernevig, and Nicolas Regnault.
\newblock Zoology of fractional chern insulators.
\newblock {\em Physical Review B—Condensed Matter and Materials Physics}, 85(7):075116, 2012.

\bibitem{regnault2011fractional}
Nicolas Regnault and B~Andrei Bernevig.
\newblock Fractional chern insulator.
\newblock {\em Physical Review X}, 1(2):021014, 2011.

\bibitem{sheng2011fractional}
DN~Sheng, Zheng-Cheng Gu, Kai Sun, and L~Sheng.
\newblock Fractional quantum hall effect in the absence of landau levels.
\newblock {\em Nature communications}, 2(1):389, 2011.

\bibitem{neupert2011fractional}
Titus Neupert, Luiz Santos, Claudio Chamon, and Christopher Mudry.
\newblock Fractional quantum hall states at zero magnetic field.
\newblock {\em Physical review letters}, 106(23):236804, 2011.

\bibitem{wang2011fractional}
Yi-Fei Wang, Zheng-Cheng Gu, Chang-De Gong, and DN~Sheng.
\newblock Fractional quantum hall effect of hard-core bosons in topological flat bands.
\newblock {\em Physical review letters}, 107(14):146803, 2011.

\bibitem{liu2013fractional}
Tianhan Liu, Cecile Repellin, B~Andrei Bernevig, and Nicolas Regnault.
\newblock Fractional chern insulators beyond laughlin states.
\newblock {\em Physical Review B—Condensed Matter and Materials Physics}, 87(20):205136, 2013.

\bibitem{kourtis2014fractional}
Stefanos Kourtis, Titus Neupert, Claudio Chamon, and Christopher Mudry.
\newblock Fractional chern insulators with strong interactions that far exceed band gaps.
\newblock {\em Physical review letters}, 112(12):126806, 2014.

\bibitem{cai2023signatures}
Jiaqi Cai, Eric Anderson, Chong Wang, Xiaowei Zhang, Xiaoyu Liu, William Holtzmann, Yinong Zhang, Fengren Fan, Takashi Taniguchi, Kenji Watanabe, et~al.
\newblock Signatures of fractional quantum anomalous hall states in twisted mote2.
\newblock {\em Nature}, 622(7981):63--68, 2023.

\bibitem{park2023observation}
Heonjoon Park, Jiaqi Cai, Eric Anderson, Yinong Zhang, Jiayi Zhu, Xiaoyu Liu, Chong Wang, William Holtzmann, Chaowei Hu, Zhaoyu Liu, et~al.
\newblock Observation of fractionally quantized anomalous hall effect.
\newblock {\em Nature}, 622(7981):74--79, 2023.

\bibitem{xu2023observation}
Fan Xu, Zheng Sun, Tongtong Jia, Chang Liu, Cheng Xu, Chushan Li, Yu~Gu, Kenji Watanabe, Takashi Taniguchi, Bingbing Tong, et~al.
\newblock Observation of integer and fractional quantum anomalous hall effects in twisted bilayer mote 2.
\newblock {\em Physical Review X}, 13(3):031037, 2023.

\bibitem{kang2024evidence}
Kaifei Kang, Bowen Shen, Yichen Qiu, Yihang Zeng, Zhengchao Xia, Kenji Watanabe, Takashi Taniguchi, Jie Shan, and Kin~Fai Mak.
\newblock Evidence of the fractional quantum spin hall effect in moir{\'e} mote2.
\newblock {\em Nature}, 628(8008):522--526, 2024.

\bibitem{zeng2023thermodynamic}
Yihang Zeng, Zhengchao Xia, Kaifei Kang, Jiacheng Zhu, Patrick Kn{\"u}ppel, Chirag Vaswani, Kenji Watanabe, Takashi Taniguchi, Kin~Fai Mak, and Jie Shan.
\newblock Thermodynamic evidence of fractional chern insulator in moir{\'e} mote2.
\newblock {\em Nature}, 622(7981):69--73, 2023.

\bibitem{lu2024fractional}
Zhengguang Lu, Tonghang Han, Yuxuan Yao, Aidan~P Reddy, Jixiang Yang, Junseok Seo, Kenji Watanabe, Takashi Taniguchi, Liang Fu, and Long Ju.
\newblock Fractional quantum anomalous hall effect in multilayer graphene.
\newblock {\em Nature}, 626(8000):759--764, 2024.

\bibitem{xie2024even}
Jian Xie, Zihao Huo, Xin Lu, Zuo Feng, Zaizhe Zhang, Wenxuan Wang, Qiu Yang, Kenji Watanabe, Takashi Taniguchi, Kaihui Liu, et~al.
\newblock Even-and odd-denominator fractional quantum anomalous hall effect in graphene moire superlattices.
\newblock {\em arXiv preprint arXiv:2405.16944}, 2024.

\bibitem{moller2015fractional}
Gunnar M{\"o}ller and Nigel~R Cooper.
\newblock Fractional chern insulators in harper-hofstadter bands with higher chern number.
\newblock {\em Physical review letters}, 115(12):126401, 2015.

\bibitem{andrews2021stability}
Bartholomew Andrews, Titus Neupert, and Gunnar M{\"o}ller.
\newblock Stability, phase transitions, and numerical breakdown of fractional chern insulators in higher chern bands of the hofstadter model.
\newblock {\em Physical Review B}, 104(12):125107, 2021.

\bibitem{liu2012fractional}
Zhao Liu, Emil~J Bergholtz, Heng Fan, and Andreas~M L{\"a}uchli.
\newblock Fractional chern insulators in topological flat bands with higher chern number.
\newblock {\em Physical review letters}, 109(18):186805, 2012.

\bibitem{wang2012fractional}
Yi-Fei Wang, Hong Yao, Chang-De Gong, and DN~Sheng.
\newblock Fractional quantum hall effect in topological flat bands with chern number two.
\newblock {\em Physical Review B}, 86(20):201101, 2012.

\bibitem{sterdyniak2013series}
Antoine Sterdyniak, Cecile Repellin, B~Andrei Bernevig, and Nicolas Regnault.
\newblock Series of abelian and non-abelian states in c> 1 fractional chern insulators.
\newblock {\em Physical Review B}, 87(20):205137, 2013.

\bibitem{wang2022hierarchy}
Jie Wang and Zhao Liu.
\newblock Hierarchy of ideal flatbands in chiral twisted multilayer graphene models.
\newblock {\em Physical Review Letters}, 128(17):176403, 2022.

\bibitem{dong2023many}
Junkai Dong, Patrick~J Ledwith, Eslam Khalaf, Jong~Yeon Lee, and Ashvin Vishwanath.
\newblock Many-body ground states from decomposition of ideal higher chern bands: Applications to chirally twisted graphene multilayers.
\newblock {\em Physical Review Research}, 5(2):023166, 2023.

\bibitem{angeli2021gamma}
Mattia Angeli and Allan~H MacDonald.
\newblock $\gamma$ valley transition metal dichalcogenide moir{\'e} bands.
\newblock {\em Proceedings of the National Academy of Sciences}, 118(10):e2021826118, 2021.

\bibitem{shen2024stabilizing}
Xiaoyang Shen, Chonghao Wang, Ruiping Guo, Zhiming Xu, Wenhui Duan, and Yong Xu.
\newblock Stabilizing fractional chern insulators via exchange interaction in moir\'e systems, 2024.

\bibitem{tong2018skyrmions}
Qingjun Tong, Fei Liu, Jiang Xiao, and Wang Yao.
\newblock Skyrmions in the moir{\'e} of van der waals 2d magnets.
\newblock {\em Nano letters}, 18(11):7194--7199, 2018.

\bibitem{xie2023evidence}
Hongchao Xie, Xiangpeng Luo, Zhipeng Ye, Zeliang Sun, Gaihua Ye, Suk~Hyun Sung, Haiwen Ge, Shaohua Yan, Yang Fu, Shangjie Tian, et~al.
\newblock Evidence of non-collinear spin texture in magnetic moir{\'e} superlattices.
\newblock {\em Nature Physics}, 19(8):1150--1155, 2023.

\bibitem{hejazi2020noncollinear}
Kasra Hejazi, Zhu-Xi Luo, and Leon Balents.
\newblock Noncollinear phases in moir{\'e} magnets.
\newblock {\em Proceedings of the National Academy of Sciences}, 117(20):10721--10726, 2020.

\bibitem{xu2022coexisting}
Yang Xu, Ariana Ray, Yu-Tsun Shao, Shengwei Jiang, Kihong Lee, Daniel Weber, Joshua~E Goldberger, Kenji Watanabe, Takashi Taniguchi, David~A Muller, et~al.
\newblock Coexisting ferromagnetic--antiferromagnetic state in twisted bilayer cri3.
\newblock {\em Nature Nanotechnology}, 17(2):143--147, 2022.

\bibitem{song2021direct}
Tiancheng Song, Qi-Chao Sun, Eric Anderson, Chong Wang, Jimin Qian, Takashi Taniguchi, Kenji Watanabe, Michael~A McGuire, Rainer St{\"o}hr, Di~Xiao, et~al.
\newblock Direct visualization of magnetic domains and moir{\'e} magnetism in twisted 2d magnets.
\newblock {\em Science}, 374(6571):1140--1144, 2021.

\bibitem{paul2023giant}
Nisarga Paul, Yang Zhang, and Liang Fu.
\newblock Giant proximity exchange and flat chern band in 2d magnet-semiconductor heterostructures.
\newblock {\em Science Advances}, 9(8):eabn1401, 2023.

\bibitem{zhang2021electronic}
Yang Zhang, Tongtong Liu, and Liang Fu.
\newblock Electronic structures, charge transfer, and charge order in twisted transition metal dichalcogenide bilayers.
\newblock {\em Physical Review B}, 103(15):155142, 2021.

\bibitem{wang2024fractional}
Chong Wang, Xiao-Wei Zhang, Xiaoyu Liu, Yuchi He, Xiaodong Xu, Ying Ran, Ting Cao, and Di~Xiao.
\newblock Fractional chern insulator in twisted bilayer mote 2.
\newblock {\em Physical Review Letters}, 132(3):036501, 2024.

\bibitem{abouelkomsan2020particle}
Ahmed Abouelkomsan, Zhao Liu, and Emil~J Bergholtz.
\newblock Particle-hole duality, emergent fermi liquids, and fractional chern insulators in moir{\'e} flatbands.
\newblock {\em Physical review letters}, 124(10):106803, 2020.

\bibitem{bernevig2012emergent}
B~Andrei Bernevig and N~Regnault.
\newblock Emergent many-body translational symmetries of abelian and non-abelian fractionally filled topological insulators.
\newblock {\em Physical Review B—Condensed Matter and Materials Physics}, 85(7):075128, 2012.

\bibitem{wang2024higher}
Chong Wang, Xiao-Wei Zhang, Xiaoyu Liu, Jie Wang, Ting Cao, and Di~Xiao.
\newblock Higher landau-level analogues and signatures of non-abelian states in twisted bilayer mote $ \_2$.
\newblock {\em arXiv preprint arXiv:2404.05697}, 2024.

\bibitem{xu2024maximally}
Cheng Xu, Jiangxu Li, Yong Xu, Zhen Bi, and Yang Zhang.
\newblock Maximally localized wannier functions, interaction models, and fractional quantum anomalous hall effect in twisted bilayer mote2.
\newblock {\em Proceedings of the National Academy of Sciences}, 121(8):e2316749121, 2024.

\bibitem{kudo2019many}
Koji Kudo, Haruki Watanabe, Toshikaze Kariyado, and Yasuhiro Hatsugai.
\newblock Many-body chern number without integration.
\newblock {\em Physical review letters}, 122(14):146601, 2019.

\bibitem{niu1985quantized}
Qian Niu, Ds~J Thouless, and Yong-Shi Wu.
\newblock Quantized hall conductance as a topological invariant.
\newblock {\em Physical Review B}, 31(6):3372, 1985.

\bibitem{bergholtz2008quantum}
EJ~Bergholtz and Anders Karlhede.
\newblock Quantum hall system in tao-thouless limit.
\newblock {\em Physical Review B}, 77(15):155308, 2008.

\bibitem{tao1983fractional}
R~Tao and DJ~Thouless.
\newblock Fractional quantization of hall conductance.
\newblock {\em Physical Review B}, 28(2):1142, 1983.

\end{thebibliography}
	
\end{document}